\documentstyle[11pt,epsf]{article}
\newcommand{\be}{\begin{equation}}
\newcommand{\ee}{\end{equation}}
\newcommand{\ben}{\begin{eqnarray}}
\newcommand{\een}{\end{eqnarray}}
\newcommand{\benn}{\begin{eqnarray*}}
\newcommand{\eenn}{\end{eqnarray*}}
\newcommand{\gb}{\bar{g}}
\newcommand{\hb}{\bar{h}}
\newcommand{\sigb}{\bar\sigma}
\newcommand{\xb}{\bar\xi}
\newcommand{\Cb}{\bar{C}}
\newcommand{\Db}{\bar{D}}
\newcommand{\Gb}{\bar{G}}
\newcommand{\mn}{{\mu\nu}}
\newcommand{\rs}{{\rho\sigma}}
\newcommand{\bD}{\mbox{\boldmath $\delta$}}
\begin{document}
\begin{titlepage}
\vspace{-20mm}
\begin{flushright}
KUCP0146
\end{flushright}
\vspace{20mm}
\centerline{\LARGE Gauge and Cutoff Function Dependence of the Ultraviolet}
\vspace{5mm}
\centerline{\LARGE Fixed Point in Quantum Gravity}
\vspace{5mm}
\centerline{{\large Wataru Souma}\footnote{\tt souma@phys.h.kyoto-u.ac.jp}}
\vspace{5mm}
\centerline{\large\it Faculty of Integrated Human Studies}
\vspace{0mm}
\centerline{\large\it Kyoto University, Kyoto 606-8501, Japan}
\vspace{50mm}
\begin{abstract}
The exact renormalization group equation for pure quantum gravity is derived
for an arbitrary gauge parameter in the space-time dimension $d=4$.
This equation is given by a non-linear functional differential equation for
the effective average action.
An action functional of the effective average action is approximated
by the same functional space of the Einstein-Hilbert action.
From this approximation, $\beta$-functions for the dimensionless
Newton constant and
cosmological constant are derived non-perturbatively.
These are used for an analysis of the phase structure and the ultraviolet
non-Gaussian fixed point of the dimensionless Newton constant.
This fixed point strongly depends on the
gauge parameter and the cutoff function.
However, this fixed point exists without
these ambiguities,
except for some gauges.
Hence, it is possible that pure quantum gravity in $d=4$ is
an asymptotically safe theory and non-perturbatively
renormalizable.
\end{abstract}
\end{titlepage}
\section{Introduction}
As is well known, quantum gravity (QG) must be treated non-perturbatively.
This is because, in the space-time dimension $d=4$,
the $L$-loop perturbative calculations in
Einstein gravity
cause divergences that are proportional
to the $L+1$ powers of the curvature tensor.
Hence, the renormalization of these terms requires infinite number of
couplings.
Thus, QG is called (perturbatively) non-renormalizable.
However, if QG is an asymptotically safe theory,
it becomes (non-perturbatively) renormalizable \cite{Weinberg}.
An asymptotically safe theory is specified by the existence of a ultraviolet
(UV) non-Gaussian fixed point (NGFP). 
The asymptotically safe nature of QG is suggested by $d=2+\epsilon$
gravity theory \cite{Weinberg,KN01}.
It is also expected that this nature will be maintained in $d=4$.
However, an ordinary perturbative $\epsilon$-expansion is
an asymptotic expansion. Thus, the large order behavior of $\epsilon$ is
not reliable.
Hence, to guarantee the existence of the UV NGFP in more higher
dimensions, the appropriate method such as the Borel resummation
must be applied.

The exact renormalization group equation (ERGE) \cite{Wilson} used
in this article
is one of the non-perturbative methods in field theories.
In addition, the applicability of the ERGE exceeds the
$\epsilon$-expansion and the $1/N$-expansion \cite{My}.
In the previous article \cite{Souma}, we used the formulation of the ERGE
for pure QG \cite{Reuter}, and clarified that
QG has the UV NGFP in $2<d\leq4$.
This result suggests that QG is an asymptotically safe theory and
(non-perturbatively) renormalizable.
However, this UV NGFP depends on the gauge parameter and the
cutoff function.
Hence, the purpose of this article is to study the effect of that
dependence to the UV NGFP.

The cutoff function dependence is an artifact problem.
The origin of this problem is stated as follows.
The ERGE is formulated as a
non-linear functional differential equation for the effective average
action. The effective average action is an infrared (IR) cutoffed 1PI
effective action, and corresponds to a coarse grained free energy.
To formulate the effective average action, we must introduce the
cutoff function. As we will discuss in Sec.~2,
the profile of the cutoff function is arbitrary.
If we can solve the ERGE without any approximations, this problem will
not appear \cite{Aoki}.
However, it is impossible to solve the ERGE without any approximations.
Hence, to reduce this equation to the calculable form, we must
truncate the functional space of the effective average action.
For this truncation, the non-linear functional differential equation
is reduced to a set of the non-linear differential equations.
This truncation causes the cutoff function dependence.

The origin of the gauge dependence in the ERGE is same as that of 
the ordinary field theory.
If we only discuss the existence of the NGFP,
this problem is not so serious.
This is because, the non-physical quantity such as the $\beta$-function and
the fixed point (FP) may depend on the gauge.
However, if this dependence causes the disappearance of the NGFP, it 
becomes problem, because the gauge dependence changes the phase structure
of the theory.
We have been believed that Abelian and Yang-Mills gauge theories have only the
GFP.
Thus, the gauge dependence of the NGFP have not been discussed.
However, the existence of the NGFP is an important point in QG.
Though we expect that the NGFP should remain for any gauges,
a strict proof is not known. 

This paper is organized as follows. In the next section, the formulation of
the ERGE for pure QG is reviewed (for details see \cite{Reuter}).
In Sec.~3, the functional space is approximated by the same space of
the Einstein-Hilbert action.
In there, the $\beta$-functions for the
dimensionless Newton constant and cosmological
constant including the constant gauge parameter are derived
by a slightly different method from other
formulations \cite{Odintsov01,Dou,Odintsov02}.
In Sec.~4, these $\beta$-functions are used to study the effect of
the gauge and cutoff function dependence to the UV NGFP.
Section 5 is devoted to summary and discussion.
\section{Formulation and approximation of the ERGE}
To derive the ERGE for pure QG,
we define the scale dependent generating functional $W_k$
of the connected Green functions.
Here, $k$ is the IR cutoff scale in the Euclidean momentum space.
Hence, $W_k$ is the IR cutoffed generating functional and defined by
\ben
e^{W_k[t,\sigb,\sigma;\beta,\tau;\gb]}
&=&\int{\cal D}h{\cal D}C{\cal D}\bar{C}\exp\left\{
-S_{\rm grav}[\gamma]-S_{\rm g.f.}[h;\gb]\right.\nonumber\\
&&-S_{\rm F.P.}[h,C,\Cb;\gb]-S_{\rm e.s.}[t,\sigb,\sigma;\beta,\tau;\gb]
\nonumber\\
&&\left.
-\Delta_kS^{\rm grav}[h;\gb]
-\Delta_kS^{\rm gh}[C,\Cb;\gb]
\right\}.\label{eq:ZW}
\een
Here, $S_{\rm grav}[\gamma]$ is a general functional of a quantum metric
$\gamma_\mn$. In the background field method, the quantum metric is
decomposed as the background metric
$\gb_\mn$ and the fluctuation from the background $h_\mn$.
The functional $S_{\rm g.f.}[h;\gb]$ in Eq.~(\ref{eq:ZW}) is the
gauge fixing term given by
\[
S_{\rm g.f.}[h;\gb]=\frac{1}{2\alpha}\int{d^dx}\sqrt{\gb}\,
\gb^\mn F_\mu F_\nu,
\]
where $\alpha$ is a gauge parameter.
In the harmonic gauge $F_\mu=0$, the explicit form of $F_\mu$ is 
given by
\[
F_\mu=\sqrt{2}\kappa{\cal F}_\mu^{\alpha\beta}[\gb]h_{\alpha\beta}
=\sqrt{2}\kappa\left(\delta_\mu^\beta\gb^{\alpha\gamma}\Db_\gamma
-\frac{1}{2}\gb^{\alpha\beta}\Db_\mu\right)h_{\alpha\beta},
\]
where $\kappa=(32\pi\Gb)^{-1/2}$ and $\Gb$ is the bare Newton constant.
The covariant derivative $\bar{D}$ is constructed from $\gb_\mn$.
In Eq.~(\ref{eq:ZW}), $S_{\rm F.P.}[h,C,\Cb;\gb]$ is the
Faddeev-Popov ghost term given by
\[
S_{\rm F.P.}[h,C,\Cb;\gb]=-\sqrt{2}\int{d^dx}\sqrt{\gb}\,
\Cb_\mu{\cal M}[\gamma;\gb]^\mu_{\;\;\;\nu} C^\nu,
\]
where $C^\mu$ and $\bar{C}_\mu$ are the ghost and anti-ghost fields
respectively.
Here,
\[
{\cal M}[\gamma;\gb]^\mu_{\;\;\;\nu}=\gb^{\mu\rho}\gb^{\sigma\lambda}
\Db_\lambda
(\gamma_{\rho\nu}D_\sigma+\gamma_{\sigma\nu}D_\rho)
-\gb^{\rho\sigma}\gb^{\mu\lambda}\Db_\lambda\gamma_{\sigma\nu}D_\rho.
\]
The functional $S_{\rm e.s.}$ in Eq.~(\ref{eq:ZW}) is the external source term
and given by
\benn
S_{\rm e.s.}[t,\sigb,\sigma;\beta,\tau;\gb]&=&
-\int{d^dx}\sqrt{\gb}\left\{t^\mn h_\mn+\sigb_\mu C^\mu+
\sigma^\mu\Cb_\mu\right.\\
&&\left.
+\beta^\mn(\bD_{\rm B}h_\mn)
+\tau_\mu(\bD_{\rm B}C^\mu)\right\}.
\eenn
Here the sources $\beta^\mn$ and $\tau_\mu$ couple to the BRST
variations of $h_\mn$ and $C^\mu$ respectively.

In Eq.~(\ref{eq:ZW}), 
$\Delta_kS^{\rm grav}[h;\gb]$ and $\Delta_kS^{\rm gh}[C,\bar{C};\gb]$
are the cutoff actions.
These terms control the propagation of fields to have the
momentum $k<p<k_0$.
Here $k_0$ is a UV cutoff scale of the theory.
Hence these are given by a quadratic form of $h_\mn$ and the ghost fields.
The explicit forms of these terms are given by
\ben
\Delta_kS^{\rm grav}[h;\gb]&=&
\frac{1}{2}\kappa^2\int d^dx\,\sqrt{\gb}\,h_\mn
(R_k^{\rm grav}[\gb])^{\mn\rho\sigma}h_{\rho\sigma},\label{eq:skgrav}\\
\Delta_kS^{\rm gh}[C,\bar{C};\gb]&=&
\sqrt{2}\int{d^dx}\sqrt{\gb}\,\Cb_\mu R_k^{\rm gh}[\gb]C^\mu,
\een
where the cutoff operator $R_k^{\rm grav}[\gb]$ and
$R_k^{\rm gh}[\gb]$ are defined by
\ben
R^{\rm grav}_k[\gb]&=&\left({\cal Z}_k^{\rm grav}\right)^{\mn\rs}
k^2R^{(0)}(-\bar{D}^2/k^2),\label{eq:cutgrav}\\
R_k^{\rm gh}[\gb]&=&Z_k^{\rm gh}k^2R^{(0)}(-\Db^2/k^2).\label{eq:cutgh}
\een
Here, $({\cal Z}_k^{\rm grav})^{\mn\rs}$ and
$Z_k^{\rm gh}$ is the renormalization factor of $h_\mn$ and the ghost field
respectively.
A convenient choice of the cutoff function $R^{(0)}(-\Db^2/k^2)$ is
\[
R^{(0)}(-\Db^2/k^2)=\frac{1+e^{-\Db^2/k^2}-e^{-\Db^2/k_0^2}}
{e^{-\Db^2/k_0^2}-e^{-\Db^2/k^2}}.\label{eq:cutfun01}
\]
If we take the limit $k_0\rightarrow\infty$, we have
\[
R^{(0)}(-\Db^2/k^2)=\frac{-\Db^2/k^2}{e^{-\Db^2/k^2}-1}.
\]
Hence, the constraints of the cutoff function $R^{(0)}(u)$ are given by
\be
\lim_{u\rightarrow0}R^{(0)}(u)=1,\hspace{5mm}
\lim_{u\rightarrow\infty}R^{(0)}(u)=0.\label{eq:cut-const}
\ee
Any functions satisfying Eq.~(\ref{eq:cut-const})
are applicable. 
Hence, the profile of the cutoff function is arbitrary.
In subsection 4.2, the effect of this ambiguity
to the UV NGFP will be considered.

The effective average action $\Gamma_k$ is defined by
\ben
\Gamma_k[\hb,\xi,\xb;\beta,\tau;\gb]&=&\int{d^dx}\sqrt{\gb}
\left\{t^\mn\hb_\mn+\sigb_\mu\xi^\mu+\sigma_\mu\xb^\mu\right\}
-W_k[t,\sigma,\sigb;\beta,\tau;\gb]
\nonumber\\
&&-\Delta_kS^{\rm grav}[\hb;\gb]-\Delta_kS^{\rm gh}[\xi,\xb;\gb].
\label{eq:eaa}
\een
Here, scale dependent classical fields are given by
\[
\hb_\mn=\frac{1}{\sqrt{\gb}}\frac{\delta W_k}{\delta t^\mn},\hspace{5mm}
\xi^\mu=\frac{1}{\sqrt{\gb}}\frac{\delta W_k}{\delta\sigb_\mu},\hspace{5mm}
\xb_\mu=\frac{1}{\sqrt{\gb}}\frac{\delta W_k}{\delta\sigma^\mu}.
\]
In addition, the classical field corresponding to $\gamma_\mn$ is
introduced as
\[
g_\mn(x)=\gb_\mn(x)+\bar{h}_\mn(x).
\]
The effective average action has two boundary conditions.
One is given by
\[
\lim_{k\rightarrow0}\Gamma_k[\bar{h},\xi,\bar\xi;\beta,\tau;\gb]
=\Gamma_{\rm 1PI}[\bar{h},\xi,\bar\xi;\beta,\tau;\gb],
\]
in the limit $k\rightarrow0$.
This is because, all quantum corrections are included
in $\Gamma_{k=0}$ in this limit.
Thus this is equivalent to the ordinary 1PI effective action.
The other is given by
\ben
\lim_{k\rightarrow\infty}\Gamma_k[g,\xi,\bar\xi;\beta,\tau;\gb]
&=&S_{\rm grav}[g]+S_{\rm g.f.}[g-\gb;\gb]
+S_{\rm F.P.}[g-\gb,\xi,\bar\xi;\gb]
\nonumber\\
&&-\int d^dx\,\sqrt{\gb}\left\{\beta^\mn(\bD_{\rm B}\hb)
+\tau_\mu(\bD_{\rm B}\xi^\mu)\right\},\label{eq:BC}
\een
in the limit $k\rightarrow\infty$, 
Here, we denote $\bar{h}_\mn$ as $g_\mn-\gb_\mn$.
Equation~(\ref{eq:BC}) means that
$\Gamma_k$ is coincide with the bare action in this limit,
since there are no quantum corrections.

If Eq.~(\ref{eq:ZW}) is differentiated with respect to $t={\rm ln}k$,
and Legendre transformed by Eq.~(\ref{eq:eaa}), we obtain
\ben
\partial_t\Gamma_k&=&\frac{1}{2}{\rm Tr}\left[
\left(\Gamma_k^{(2)}+\kappa^2R_k^{\rm grav}\right)^{-1}_{\bar{h}\bar{h}}
(\partial_t\kappa^2R_k^{\rm grav})^{\mn\rho\sigma}\right]\nonumber\\
&&-\frac{1}{2}{\rm Tr}\left[\left\{
\left(\Gamma_k^{(2)}\!+\!\sqrt{2}R_k^{\rm gh}\right)^{-1}_{\bar{\xi}\xi}
\!\!-\left(\Gamma_k^{(2)}\!+\!\sqrt{2}R_k^{\rm gh}\right)^{-1}_{\xi\bar{\xi}}
\right\}(\partial_t\sqrt{2}R_k^{\rm gh})
\right].\label{eq:FE01}
\een
This is the ERGE for pure QG.
Here $\Gamma_k^{(2)}$ is the Hessian
of $\Gamma_k$ with respect to the subscript.

Though Eq.~(\ref{eq:FE01}) is non-perturbatively exact, the manifest
BRST invariance is broken by the explicit IR cutoff.
To see this case more detail, we consider the Ward-Takahashi (WT)
identity.
Now, the WT identity is given by
\[
0=\left<\bD_B S_{\rm e.s.}+\bD_B\Delta_kS^{\rm grav}
+\bD_B\Delta_kS^{\rm gh}\right>.
\]
If this is written in terms of the effective average action, we have
\be
\int d^dx\frac{1}{\sqrt{\gb}}\left\{
\frac{\delta\Gamma'_k}{\delta\bar{h}_\mn}
\frac{\delta\Gamma'_k}{\delta\beta^\mn}
+\frac{\delta\Gamma'_k}{\delta\xi^\mu}
\frac{\delta\Gamma'_k}{\delta\tau_\mu}
\right\}
=Y_k\left(R_k^{\rm grav}[\gb],R_k^{\rm gh}[\gb]\right)
\label{eq:mwti},
\ee
where $\Gamma'_k=\Gamma_k-S_{\rm g.f.}$ is introduced.
In the usual field theories, the RHS of Eq.~(\ref{eq:mwti}) equals to zero.
In the present case, the existence of the cutoff action makes it
proportional to cutoff operators.
The RHS of Eq.~(\ref{eq:mwti}) goes to zero in the
limit $k\rightarrow0$, because
cutoff operators goes to zero in this limit.
Hence, an usual field theory is recovered.
However, $Y_k$ does not disappear in the intermediate scale $k$.
Thus, the BRST symmetry is broken in this scale.

Now, to get the BRST invariant RG flows,
we approximate the functional space.
As a first step approximation, we neglect
the evolution of the ghost action and the external source fields.
From this approximation, the effective averaged
action is expected as
\ben
\Gamma_k[g,\xi,\bar\xi;\beta,\tau;\gb]
&=&\bar\Gamma_k[g]+\hat\Gamma_k[g,\gb]+S_{\rm g.f.}[g-\gb;\gb]
+S_{\rm F.P.}[g-\gb,\xi,\bar\xi;\gb]\nonumber\\
&&-\int d^dx\,\sqrt{\gb}\left\{\beta^\mn(\bD_{\rm B}\hb_\mn)
+\tau_\mu(\bD_{\rm B}\xi^\mu)\right\}.\label{eq:exgamma}
\een
Here $S_{\rm g.f.}$ and $S_{\rm F.P.}$ have the same form as in 
the bare action.
The coupling to the BRST variations also has the same form
as in the bare action.
The remaining term is decomposed into $\bar\Gamma_k[g]$ and $\hat\Gamma_k[g,\gb]$.
Here, $\hat\Gamma_k[g,\gb]$ contains
the deviations for $g_\mn\neq\gb_\mn$, and 
satisfies $\hat\Gamma_k[g,\gb=g]=0$. 
The approximated effective average action given by Eq.~(\ref{eq:exgamma})
satisfies the boundary condition of Eq. (\ref{eq:BC}),
if these terms satisfy
\[
\lim_{k\rightarrow\infty}\bar\Gamma_k[g]=S_{\rm grav}[g],\hskip 5mm
\lim_{k\rightarrow\infty}\hat\Gamma_k[g,\gb]=0.
\]
These conditions suggest that setting $\hat\Gamma_k[g,\gb]=0$ for all $k$ is
the candidate to get the BRST invariant
RG flows. 
Substituting Eq.~(\ref{eq:exgamma}) into Eq.~(\ref{eq:mwti}), we obtain
\be
\int d^dx{\cal L}_\xi g_\mn\frac{\delta\hat\Gamma_k[g,\gb]}{\delta g_\mn(x)}
=-Y_k\left(R_k^{\rm grav}[\gb],R_k^{\rm gh}[\gb]\right), \label{eq:MODWT}
\ee
where ${\cal L}_\xi$ means the Lie derivative with respect to $\xi^\mu$.
The RHS of  Eq.~(\ref{eq:MODWT}) is regarded as the higher loop corrections
if it is evaluated perturbatively (for details see \cite{Reuter}).
Hence to neglect $Y_k$ is acceptable in the first approximation.
This is consistent with setting $\hat\Gamma_k=0$ for all scales.
These approximation means that the RG flows are projected on $g_\mn=\gb_\mn$
in all scales.
In the background spaces the BRST invariance is preserved.
Hence, the projected RG flows moving in the background space
are regarded as the BRST invariant.

If Eq.~(\ref{eq:exgamma}) is inserted into Eq.~(\ref{eq:FE01}),
the approximated ERGE becomes
\be
\partial_t\Gamma_k[g;\gb]={\cal S}_k^{\rm grav}-{\cal S}_k^{\rm gh},
\label{eq:FE02}
\ee
where,
\be
\Gamma_k[g;\gb]=\bar\Gamma_k[g]+\hat\Gamma[g;\gb]+S_{\rm g.f.}[g-\gb;\gb].
\label{eq:prog}
\ee
This has the boundary condition
\[
\lim_{k\rightarrow\infty}\Gamma_k[g;\gb]=S_{\rm grav}[\gb]+
S_{\rm g.f.}[g-\gb;\gb].
\]
In Eq.~(\ref{eq:FE02}), ${\cal S}_k^{\rm grav}$ and ${\cal S}_k^{\rm gh}$
correspond to the gravitational sector and the ghost sector respectively, and
are given by
\ben
{\cal S}_k^{\rm grav}&=&\frac{1}{2}{\rm Tr}\left[
\left(\kappa^{-2}\Gamma_k^{(2)}[g;\gb]+R_k^{\rm grav}[\gb]\right)^{-1}
(\partial_t R_k^{\rm grav}[\gb])\right],\label{eq:skgravv}\\
{\cal S}_k^{\rm gh}&=&-{\rm Tr}\left[
\left(-{\cal M}[g;\gb]+R_k^{\rm gh}[\gb]\right)^{-1}
(\partial_tR_k^{\rm gh}[\gb])\right].\label{ea:skgh}
\een
In Eq.~(\ref{eq:skgravv}), $\Gamma_k^{(2)}[g;\gb]$ represents the Hessian
of $\Gamma_k[g;\gb]$ with respect to $g_\mn$ at fixed $\gb_\mn$.

\section{Einstein-Hilbert truncation}

In below we consider the case $d=4$.
Now, to make problems easier, we truncate
the functional space of $\Gamma_k[g;\gb]$.
The most naive truncation is to take the functional space
as a same space of the Einstein-Hilbert action. Thus the bare action is
given by
\[
S_{\rm grav}[g]=\frac{1}{16\pi\bar{G}}\int d^4x\,\sqrt{g}\left\{
-R(g)+2\bar{\lambda}\right\},
\]
where $\bar\lambda$ is the bare cosmological
constant. Now, we define the scale dependent couplings as
\[
\bar{G}\rightarrow G_k=Z_{Nk}^{-1}\bar{G},
\hskip 5mm \bar\lambda\rightarrow\bar\lambda_k,
\hskip 5mm \alpha\rightarrow\alpha_k=Z_{Nk}^{-1}\alpha.
\]
Hence, from Eq.~(\ref{eq:prog}), $\Gamma_k[g;\gb]$ is expected as
\ben
\Gamma_k[g;\gb]&=&2\kappa^2Z_{Nk}\int d^4x\,\sqrt{g}
\left\{-R(g)+2\bar\lambda_k\right\}\nonumber\\
&&
+\frac{\kappa^2}{\alpha}Z_{Nk}
\int d^4x\,\sqrt{\gb}\,\gb^\mn
({\cal F}^{\alpha\beta}_\mu g_{\alpha\beta}) 
({\cal F}^{\rho\sigma}_\nu g_{\rho\sigma}).\label{eq:EAA}
\een
Here $Z_{Nk}$ is a renormalization factor.
If Eq.~(\ref{eq:EAA}) is differentiated with respect to $t$
and projected on $g_\mn=\gb_\mn$, we have
\be
\partial_t\Gamma_k[g;g]=2\kappa^2\int d^4x\,\sqrt{g}\left[
-R(g)\partial_tZ_{Nk}+2\partial_t(Z_{Nk}\bar\lambda_k)\right].
\label{eq:LHS}
\ee
This is the LHS of Eq.~(\ref{eq:FE02}).
Though the differentiation with respect to $t$ brings the term 
that is proportional to the ghost action, this term
disappears on $g_\mn=\gb_\mn$ because 
${\cal F}_\mu^{\alpha\beta}g_{\alpha\beta}|_{g=\gb}=0$.
Thus the gauge parameter is treated as a constant.
This is a problem of the present formulation.

Next step is to get the RHS of Eq.~(\ref{eq:FE02}).
Firstly, we calculate ${\cal S}_k^{\rm grav}$ given by
Eq.~(\ref{eq:skgravv}).
Now, we can naively write Eq.~(\ref{eq:skgravv}) as
\ben
{\cal S}_k^{\rm grav}&=&\frac{1}{2}\partial_t{\rm Tr}\ln
\left(\kappa^{-2}\Gamma_k^{(2)}[g;\gb]+R_k^{\rm grav}[\gb]\right)\nonumber\\
&&-\frac{1}{2}{\rm Tr}\left[
\left(\kappa^{-2}\Gamma_k^{(2)}[g;\gb]+R_k^{\rm grav}[\gb]\right)^{-1}
\left(\kappa^{-2}\partial_t\Gamma_k^{(2)}[g,\gb]\right)
\right].\label{eq:RHS1}
\een
The FP action $\Gamma^*[g;\gb]$ satisfies
$\partial_t\Gamma^*[g;\gb]=0$. Hence, if we are interested in only the
FP solution, we can neglect the second term in the RHS of Eq.~(\ref{eq:RHS1}).
Thus, we have
\be
\label{eq:RHS2}
{\cal S}_k^{\rm grav}=-\partial_t\ln I_k^{\rm grav}[\gb],
\ee
where,
\be
\label{eq:RHS3}
I_k^{\rm grav}[\gb]=
\int{\cal D}\bar{h}_\mn\exp\left\{-\Gamma_k^{\rm quad}[\bar{h};\gb]
-\Delta_kS^{\rm grav}[\bar{h};\gb]
\right\}.
\ee
Here, $\Delta_kS^{\rm grav}$ is given similarly to Eq.~(\ref{eq:skgrav})
except for the change of the field: $h_\mn\rightarrow\bar{h}_\mn$, and
$\Gamma_k^{\rm quad}[\bar{h},\gb]$ is defined by
\[
\kappa^{-2}\Gamma_k[g=\gb+\bar{h};\gb]
=\Gamma_k[\gb]+O(\bar{h})+\Gamma_k^{\rm quad}[\bar{h};\gb]
+O(\bar{h}^3).
\]
The explicit form is given by
\ben
\Gamma_k^{\rm quad}[\bar{h};\gb]&=&
Z_{Nk}\int{d^4x}\sqrt{\gb}\;
\bar{h}_\mn\left[-K^\mn_{\;\;\;\;\rs}\bar{D}^2
+U^\mn_{\;\;\;\;\rs}\right]\bar{h}^\rs\nonumber\\
&&-\left(1-\frac{1}{\alpha}\right)Z_{Nk}\int{d^4x}\sqrt{\gb}\;\gb^\mn
\left({\cal F}_\mu^{\alpha\beta}\bar{h}_{\alpha\beta}\right)
\left({\cal F}_\nu^\rs\bar{h}_\rs\right),
\label{eq:quad1}
\een
where,
\benn
K^\mn_{\;\;\;\;\rs}&=&\frac{1}{4}\left(\delta^\mu_\rho\delta^\nu_\sigma
+\delta^\mu_\sigma\delta^\nu_\rho-\gb^\mn\gb_\rs
\right),\\
U^\mn_{\;\;\;\;\rs}&=&K^\mn_{\;\;\;\;\rs}\left(\bar{R}-2\bar\lambda_k\right)
+\frac{1}{2}\left(
\gb^\mn\bar{R}_\rs+\gb_\rs\bar{R}^\mn
\right)-\frac{1}{2}\left(
\bar{R}^{\nu\;\;\mu}_{\;\;\rho\;\;\sigma}
+\bar{R}^{\nu\;\;\mu}_{\;\;\sigma\;\;\rho}
\right)
\\
&&-\frac{1}{4}\left(
\delta^\mu_\rho\bar{R}^\nu_{\;\;\sigma}
+\delta^\mu_\sigma\bar{R}^\nu_{\;\;\rho}
+\delta^\nu_\rho\bar{R}^\mu_{\;\;\sigma}
+\delta^\nu_\sigma\bar{R}^\mu_{\;\;\rho}\right).
\eenn
Hence, if the one-loop effective action $I_k^{\rm grav}$ is
calculated, we can get ${\cal S}_k^{\rm grav}$ from Eq.~(\ref{eq:RHS2}).
Now, to calculate the one-loop effective action,
we decompose $\bar{h}_\mn$ as
\be
\bar{h}_\mn=\hat{h}_\mn^\perp
+\left(\bar{D}_\mu\xi_\nu^\perp+\bar{D}_\nu\xi_\mu^\perp\right)
+\left(\bar{D}_\mu\bar{D}_\nu-\frac{1}{4}\gb_\mn\bar{D}^2\right)\sigma
+\frac{1}{4}\gb_\mn\phi
\label{eq:decomp},
\ee
where $\phi=\gb^\mn\bar{h}_\mn$, and $\xi^\perp_\mu$ satisfies
$\bar{D}^\mu\xi_\mu^\perp=0$ \cite{Odintsov01,Duff}.
In addition, $\hat{h}_\mn^\perp$ satisfies
$\bar{D}^\mu\hat{h}_\mn^\perp=0$ and $\gb^\mn\hat{h}_\mn^\perp=0$.
The resulting measure and jacobian are given by
\ben
{\cal D}\bar{h}_\mn&\rightarrow&{\cal D}\hat{h}^\perp_\mn{\cal D}
\xi^\perp_\lambda
{\cal D}\sigma\left[\det J\right]^{1/2},\label{eq:measure}\\
J&=&\Delta_1\left(-\frac{1}{4}\bar{R}\right)\otimes
\Delta_0\left(-\frac{1}{3}\bar{R}\right)\otimes
\Delta_0\left(0\right)\label{eq:jac}.
\een

Furthermore, we take the background as the maximally symmetric.
In this background, the Riemann tensor $\bar{R}_{\mn\rs}$ 
and the Ricchi tensor $\bar{R}_\mn$ are given by
\benn
\bar{R}_{\mn\rs}&=&\frac{1}{12}\left(\gb_{\mu\rho}\gb_{\nu\sigma}
-\gb_{\mu\sigma}\gb_{\nu\rho}\right)\bar{R},\\
\bar{R}_\mn&=&\frac{1}{4}\gb_\mn\bar{R}.
\eenn
Here, the scalar curvature $\bar{R}$ characterizes the space.
Now, we introduce the constrained operators as
\benn
\Delta_0(X)\phi&=&\left(-\bar{D}^2+X\right)\phi,\\
\Delta_{1\mn}(X)\xi^\perp_\nu&=&
\left(-\bar{D}^2_\mn+\gb_\mn X\right)\xi^\perp_\nu,\\
\Delta_{2\alpha\beta}^\mn(X)\hat{h}^\perp_\mn&=&
\left(-\bar{D}_{\alpha\beta}^{2\mn}+\delta^\mu_\alpha\delta_\beta^\nu X\right)
\hat{h}^\perp_\mn.
\eenn
From these manipulations, Eq.~(\ref{eq:quad1}) becomes
\ben
\Gamma_k^{\rm quad}[\bar{h};\gb]&=&
Z_{Nk}\int{d^4x}\sqrt{\gb}\left\{
\frac{1}{2}\hat{h}_\mn^\perp\Delta_2\left(
-2\bar\lambda_k+\frac{2}{3}\bar{R}\right)
\hat{h}^{\perp\mn}\right.\nonumber\\
&&+\frac{1}{\alpha}\hat\xi_\mu^\perp\Delta_1\left(
-2\alpha\bar\lambda_k+\frac{2\alpha-1}{4}\bar{R}
\right)\hat\xi^{\perp\mu}\nonumber\\
&&-\frac{3(\alpha-3)}{16\alpha}\left[
\hat\sigma\Delta_0\left(+\frac{4\alpha}{\alpha-3}\bar\lambda_k
-\frac{\alpha-1}{\alpha-3}\bar{R}\right)
\hat\sigma\right.\nonumber\\
&&
+\frac{2(\alpha-1)}{\alpha-3}\hat\sigma
\sqrt{\Delta_0\left(0\right)}
\sqrt{\Delta_0\left(-\bar{R}/3\right)}\phi\nonumber\\
&&\left.\left.
+\frac{3\alpha-1}{3(\alpha-3)}\phi\Delta_0\left(
-\frac{2\alpha}{3\alpha-1}\bar\lambda_k\right)\phi
\right]\right\}.\label{eq:quad2}
\een
Here, we introduced
\be
\hat\xi^\perp_\mu=
\sqrt{\Delta_1\left(-\frac{1}{4}\bar{R}\right)}\;\xi^\perp_\mu,\hspace{5mm}
\hat\sigma=
\sqrt{\Delta_0(0)}\sqrt{\Delta_0\left(-\frac{1}{3}\bar{R}\right)}\;\sigma.
\label{eq:change}
\ee
The changes of variables in Eq.~(\ref{eq:change}) cancel
the jacobian in Eqs.~(\ref{eq:measure}) and (\ref{eq:jac}).

To get the one-loop effective action given by Eq.~(\ref{eq:RHS3}), we
decompose $\bar{h}_\mn$ in $\Delta_kS^{\rm grav}[\bar{h};\gb]$.
Up to now, the form of ${\cal Z}_k^{\rm grav}$ in Eq.~(\ref{eq:cutgrav})
is not specified.
Now, we consider ${\cal Z}_k^{\rm grav}$ as the non-trivial projection
operator which makes $\Delta_kS^{\rm grav}[\bar{h};\gb]$ to
\ben
\Delta_kS^{\rm grav}[\bar{h},\gb]&=&Z_{Nk}\int{d^4x}\sqrt{\gb}\left[
\frac{1}{2}\hat{h}^\perp_\mn\left(k^2R^{(0)}\right)
\hat{h}^{\perp\mn}
+\frac{1}{\alpha}\hat\xi^\perp_\mu\left(k^2R^{(0)}\right)
\hat\xi^{\perp\mu}\right.\nonumber\\
&&-\frac{3(\alpha-3)}{16\alpha}\left\{
\hat\sigma\left(k^2R^{(0)}\right)\hat\sigma
+\frac{3\alpha-1}{3(\alpha-1)}\phi\left(k^2R^{(0)}\right)\phi\right.
\nonumber\\
&&
+\frac{2(\alpha-1)}{\alpha-3}\hat\sigma
\left(
\sqrt{\Delta_0\left(k^2R^{(0)}\right)}
\sqrt{\Delta_0\left(k^2R^{(0)}-\bar{R}/3\right)}
\right.\nonumber\\
&&\left.\left.\left.
-\sqrt{\Delta_0(0)}\sqrt{\Delta_0(-\bar{R}/3)}
\right)\phi
\right\}\right].\label{eq:skgravvv}
\een
Here, we introduced the shorthand
notation $R^{(0)}(-D^2/k^2)\equiv R^{(0)}$ \cite{Dou}.
Hence, from Eqs.~(\ref{eq:quad2}) and (\ref{eq:skgravvv}), we can calculate
$I_k^{\rm grav}$ by the Gaussian integral.
However, that result includes the additional zero-modes
introduced by
the decomposition in Eq.~(\ref{eq:decomp}).
Then to remove these modes \cite{Odintsov01,Duff},
we introduce unconstrained
operators as
\benn
\det\Delta_{\rm S}\left(X\right)&=&\det\Delta_0\left(X\right),\\
\det\Delta_{\rm V}\left(X\right)&=&\det\Delta_1\left(X\right)
\det\Delta_0\left(X-\frac{1}{4}\bar{R}\right),\\
\det\Delta_{\rm T}\left(X\right)&=&\det\Delta_2\left(X\right)
\det\Delta_1\left(X-\frac{5}{12}\bar{R}\right)
\det\Delta_0\left(X-\frac{2}{3}\bar{R}\right).
\eenn
Hence, we have
\benn
I_k^{\rm grav}[\gb]&=&\left[\det Z_{Nk}\Delta_{\rm T}
\left(k^2R^{(0)}
-2\bar\lambda_k+\frac{2}{3}\bar{R}\right)\right]^{-\frac{1}{2}}
\\
&&\cdot\left[\det Z_{Nk}\Delta_{\rm V}\left(k^2R^{(0)}
-2\alpha\bar\lambda_k+\frac{2\alpha-1}{4}\bar{R}
\right)\right]^{-\frac{1}{2}}
\\
&&\cdot\left[\det Z_{Nk}\Delta_{\rm V}\left(k^2R^{(0)}
-2\bar\lambda_k+\frac{1}{4}\bar{R}
\right)\right]^{\frac{1}{2}}\\
&&\cdot\left[\det Z_{Nk}\Delta_{\rm S}\left(k^2R^{(0)}
-2\bar\lambda_k\right)\right]^{-\frac{1}{2}}.
\eenn
Therefore, from Eq.~(\ref{eq:RHS3}),
\ben
{\cal S}_k^{\rm grav}&=&
\frac{1}{2}{\rm Tr_T}\left[
{\cal N}\left({\cal A}+\frac{2}{3}R\right)^{-1}\right]
+\frac{1}{2}{\rm Tr_V}\left[
{\cal N}\left({\cal A}_\alpha+\frac{2\alpha-1}{4}R\right)^{-1}\right]\nonumber\\
&&-\frac{1}{2}{\rm Tr_V}\left[{\cal N}
\left({\cal A}+\frac{1}{4}R\right)^{-1}\right]
+\frac{1}{2}{\rm Tr_S}\left[{\cal N}{\cal A}^{-1}\right]\label{eq:right},
\een
where, ${\cal A},\;{\cal A}_\alpha$ and ${\cal N}$ are given by
\benn
{\cal A}&=&-D^2+k^2R^{(0)}(-D^2/k^2)-2\bar\lambda_k,\\
{\cal A}_\alpha&=&-D^2+k^2R^{(0)}(-D^2/k^2)-2\alpha\bar\lambda_k,\\
{\cal N}&=&\left(1-\frac{\eta}{2}\right)k^2R^{(0)}(-D^2/k^2)
+D^2R^{(0)\prime}(-D^2/k^2).
\eenn
Here, the prime means the differentiation with respect to the argument.
The anomalous dimension $\eta$ is defined by
$\eta=-\partial_t\ln Z_{Nk}$.
In Eq.~(\ref{eq:right}), $g_\mn=\gb_\mn$ is taken.
In below, we omit the bars from the metric and the scalar curvature.

The remaining term of the RHS of Eq.~(\ref{eq:FE02}) is ${\cal S}_k^{\rm gh}$.
For the Faddeev-Popov ghost term, we do not take into account the renormalization
of these field. Hence, $R_k^{\rm gh}=k^2R^{(0)}(-D^2/k^2)$.
In the present background, ${\cal M}=-D^2-R/4$.
Therefore,
\be
{\cal S}_k^{\rm gh}=
-{\rm Tr_V}\left[{\cal N}_0\left({\cal A}_0-\frac{1}{4}R\right)^{-1}\right].
\label{eq:gammabt}
\ee
Here, ${\cal A}_0$ and ${\cal N}_0$ in Eq.~(\ref{eq:gammabt}) are
defined similarly to ${\cal A}$ and ${\cal N}$ except for $\bar\lambda_k=0$
and $\eta=0$.
In below, we denote the RHS of Eq.~(\ref{eq:FE02}) as
${\cal S}_{\rm R}={\cal S}_k^{\rm grav}+{\cal S}_k^{\rm gh}$.

Now to get the coefficients of $\sqrt{g}$ and $\sqrt{g}R$,
we expand ${\cal S}_{\rm R}$ in terms of the scalar curvature $R$,
\ben
{\cal S}_{\rm R}
&=&\frac{1}{2}{\rm Tr_T}\left[{\cal N}{\cal A}^{-1}\right]
+\frac{1}{2}{\rm Tr_V}\left[{\cal N}{\cal A}_\alpha^{-1}\right]
-\frac{1}{2}{\rm Tr_V}\left[{\cal N}{\cal A}^{-1}\right]\nonumber\\
&&+\frac{1}{2}{\rm Tr_S}\left[{\cal N}{\cal A}^{-1}\right]
-{\rm Tr_V}\left[{\cal N}_0{\cal A}_0^{-1}\right]\nonumber\\
&&
-R\left\{
\frac{1}{3}{\rm Tr_T}\left[{\cal N}{\cal A}^{-2}\right]
+\frac{2\alpha-1}{8}{\rm Tr_V}\left[{\cal N}{\cal A}_\alpha^{-2}\right]
\right.\nonumber\\
&&\left.-\frac{1}{8}{\rm Tr_V}\left[{\cal N}{\cal A}^{-2}\right]
+\frac{1}{4}{\rm Tr_V}\left[{\cal N}_0{\cal A}_0^{-2}\right]
\right\}+O(R^2).
\label{eq:QUAD05}
\een
Furthermore, to calculate the traces in Eq.~(\ref{eq:QUAD05}),
we use the heat kernel expansion:
\ben
{\rm Tr}_j\left[W(-D^2)\right]&=&(4\pi)^{-2}{\rm tr}_j(I)\left\{
Q_2[W]\int{d^4x}\sqrt{g}\right.\nonumber\\
&&\left.+\frac{1}{6}Q_1[W]\int{d^4x}\sqrt{g}R+O(R^2)
\right\},\label{eq:HKE}
\een
where, $I$ is a unit matrix and $j={\rm T,V,S}$ mean the tensor, vector
and scalar respectively.
In Eq.~(\ref{eq:HKE}), ${\rm tr}_j(I)$ simply counts the number of
independent degrees of freedom of these quantity:
\[
{\rm tr_T}(I)=9,\hspace{5mm}
{\rm tr_V}(I)=4,\hspace{5mm}
{\rm tr_S}(I)=1.
\]
In Eq.~(\ref{eq:HKE}), $Q_i[W],\;(i=1,2)$
is the Mellin transform of $W$, and given by
\benn
Q_0[W]&=&W(0),\\
Q_n[W]&=&\frac{1}{\Gamma(n)}\int_0^\infty dz z^{n-1}W(z),\hspace{3mm}(n>0).
\eenn
Therefore, if we insert Eq.~(\ref{eq:HKE}) into Eq.~(\ref{eq:QUAD05}),
and compare it
with Eq.~(\ref{eq:LHS}),
\ben
\partial_t\left(Z_{Nk}\bar\lambda_k\right)&=&
\frac{1}{4\kappa^2}\frac{k^4}{(4\pi^2)}\left[
6\Phi_2^1(-2\bar\lambda_k/k^2)+4\Phi_2^1(-2\alpha\bar\lambda_k/k^2)
\right.\nonumber\\
&&\left.
-\eta\left\{3\widetilde\Phi_2^1(-2\bar\lambda_k/k^2)
+2\widetilde\Phi_2^1(-2\alpha\bar\lambda_k/k^2)\right\}
\right],\label{eq:DEQ1}\\
\partial_tZ_{Nk}&=&
-\frac{1}{12\kappa^2}\frac{k^2}{(4\pi)^2}\left[
-30\Phi_2^2(-2\bar\lambda_k/k^2)
-6(2\alpha-1)\Phi_2^2(-2\alpha\bar\lambda_k/k^2)\right.\nonumber\\
&&+6\Phi_1^1(-2\bar\lambda_k/k^2)
+4\Phi_1^1(-2\alpha\bar\lambda_k/k^2)
-12\Phi_2^2(0)-8\Phi_1^1(0)\nonumber\\
&&-\eta\left\{
-15\widetilde\Phi_2^2(-2\bar\lambda_k/k^2)
-3(2\alpha-1)\widetilde\Phi_2^2(-2\alpha\bar\lambda_k/k^2)\right.\nonumber\\
&&\left.\left.
+3\widetilde\Phi_1^1(-2\bar\lambda_k/k^2)
+\widetilde\Phi_1^1(-2\alpha\bar\lambda_k/k^2)
\right\}
\right].
\label{eq:DEQ2}
\een
Here 
$\Phi^p_n(w)$ and $\widetilde\Phi^p_n(w)$ are
concerning with the integrals of the cutoff function, and defined by
\benn
\Phi_n^p(w)&=&\frac{1}{\Gamma(n)}\int_0^\infty dx x^{n-1}
\frac{R^{(0)}(x)-xR^{(0)\prime}(x)}{\left[x+R^{(0)}(x)+w\right]^p},\\
\widetilde\Phi_n^p(w)&=&\frac{1}{\Gamma(n)}\int_0^\infty dx x^{n-1}
\frac{R^{(0)}(x)}{\left[x+R^{(0)}(x)+w\right]^p}.
\eenn
Here, the cutoff function $R^{(0)}(x)$ satisfies constraints
given by Eq.~(\ref{eq:cut-const}).
In \cite{Reuter}, this is given by
\be
R^{(0)}(x)=\frac{x}{\exp\left(x\right)-1}.\label{eq:CUTF}
\ee

Now, we introduce the dimensionless Newton constant $g_k$ and cosmological
constant $\lambda_k$,
\[
g_k=k^2G_k=k^2Z_{Nk}^{-1}\bar{G},\hskip 5mm
\lambda_k=k^{-2}\bar\lambda_k.
\]
If Eqs.~(\ref{eq:DEQ1}) and (\ref{eq:DEQ2})
are expressed in terms of dimensionless couplings, we obtain
\ben
&&\beta_g=\partial_t g_k=(2+\eta)g_k,\label{eq:BETAG}\\
&&\beta_\lambda=\partial_t\lambda_k=-(2-\eta)\lambda_k
+g_kB_3(\lambda_k,\alpha).\label{eq:BETAL}
\een
In this case, the anomalous dimension is given by
\be
\eta=
\frac{g_kB_1(\lambda_k,\alpha)}{1-g_kB_2(\lambda_k,\alpha)}.
\label{eq:eta}
\ee
Now, $B_i(\lambda_k,\alpha),\;(i=1,2,3)$ in
Eqs.~(\ref{eq:BETAL}) and (\ref{eq:eta}) is defined by
\benn
B_1(\lambda_k,\alpha)&=&\frac{1}{3\pi}\left[
3\Phi_1^1(-2\lambda_k)
+2\Phi_1^1(-2\alpha\lambda_k)
-4\Phi_1^1(0)
\right.\nonumber\\
&&\left.
-15\Phi_2^2(-2\lambda_k)
-3(2\alpha-1)\Phi_2^2(-2\alpha\lambda_k)
-6\Phi_2^2(0)
\right],\\
B_2(\lambda_k,\alpha)&=&-\frac{1}{6\pi}\left[
3\widetilde\Phi_1^1(-2\lambda_k)
+2\widetilde\Phi_1^1(-2\alpha\lambda_k)
\right.\\
&&\left.
-15\widetilde\Phi_2^2(-2\lambda_k)
-3(2\alpha-1)\widetilde\Phi_2^2(-2\alpha\lambda_k)
\right],\\
B_3(\lambda_k,\alpha)&=&\frac{1}{2\pi}
\left[6\Phi_2^1(-2\lambda_k)
+4\Phi_2^1(-2\alpha\lambda_k)
-8\Phi_2^1(0)
\right.\\
&&\left.
-\eta\left\{
3\widetilde\Phi_2^1(-2\lambda_k)
+2\widetilde\Phi_2^1(-2\alpha\lambda_k)\right\}\right].
\eenn
These results are same as that of \cite{Odintsov01}. If $\alpha=1$, these
reproduce the results of \cite{Reuter}.

\section{The UV NGFP of QG}
\subsection{Gauge dependence}
In below, we ignore the dimensionless cosmological constant.
This approximation is
reliable when the dimensionless cosmological constant is much smaller
than the IR
cutoff scale: $\lambda_k\ll k$.
This approximation is applicable if
we are interested in the local structure of the Universe.

On the FP, the scale invariance is preserved.
Hence, if we denote $g^*$ as the FP of the dimensionless Newton constant,
this satisfies
\be
0=(2+\eta^*)g^*.\label{eq:FPGcond}
\ee
From Eq.~(\ref{eq:FPGcond}), it is recognized that the candidates
of the FP are
$g^*=0$ and $\eta^*=-2$.
Here the former is the GFP and exists independently of $\alpha$.
The latter is the candidate of the NGFP.
The condition $\eta^*=-2$ reads to
\be
g^*=\frac{-2}{B_1(\alpha)-2B_2(\alpha)}.\label{eq:FPG}
\ee
Here, $B_i(\alpha),\;(i=1,2)$ is given by
\benn
B_1(\alpha)&=&\frac{1}{3\pi}\left[\Phi_1^1(0)-6(\alpha+3)\Phi_2^2(0)\right],\\
B_2(\alpha)&=&-\frac{1}{6\pi}\left[5\widetilde\Phi_1^1(0)
-6(\alpha+2)\widetilde\Phi_2^2(0)\right].
\eenn

Now, to calculate $\Phi^i_i(0)$ and
$\widetilde\Phi^i_i(0)\;(i=1,2)$,
we use the cutoff function given by Eq.~(\ref{eq:CUTF}).
Hence we have
\[
\Phi^1_1(0)=\frac{\pi^2}{6},\hspace{5mm}
\Phi^2_2(0)=1,\hspace{5mm}
\widetilde\Phi^1_1(0)=1,\hspace{5mm}
\widetilde\Phi^2_2(0)=\frac{1}{2}.
\]
Therefore, Eq.~(\ref{eq:FPG}) becomes
\be
g^*=\frac{2\pi}{3}\left(\alpha-\frac{\pi^2-114}{54}\right)^{-1}
\label{eq:FPG4}.
\ee
\begin{figure}[t]
\epsfxsize=0.49\textwidth
\begin{center}
\leavevmode
\epsffile{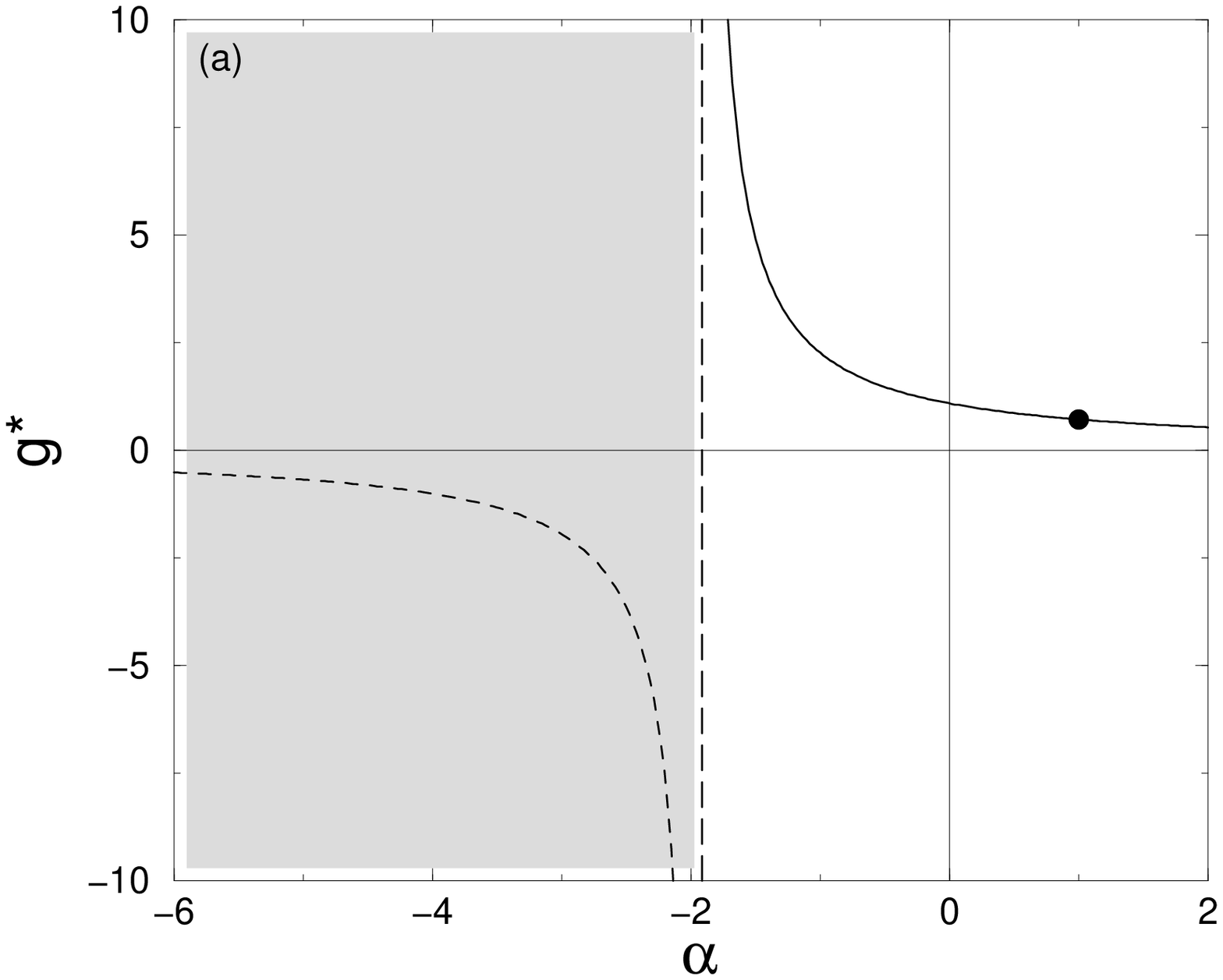}
\epsfxsize=0.49\textwidth
\epsffile{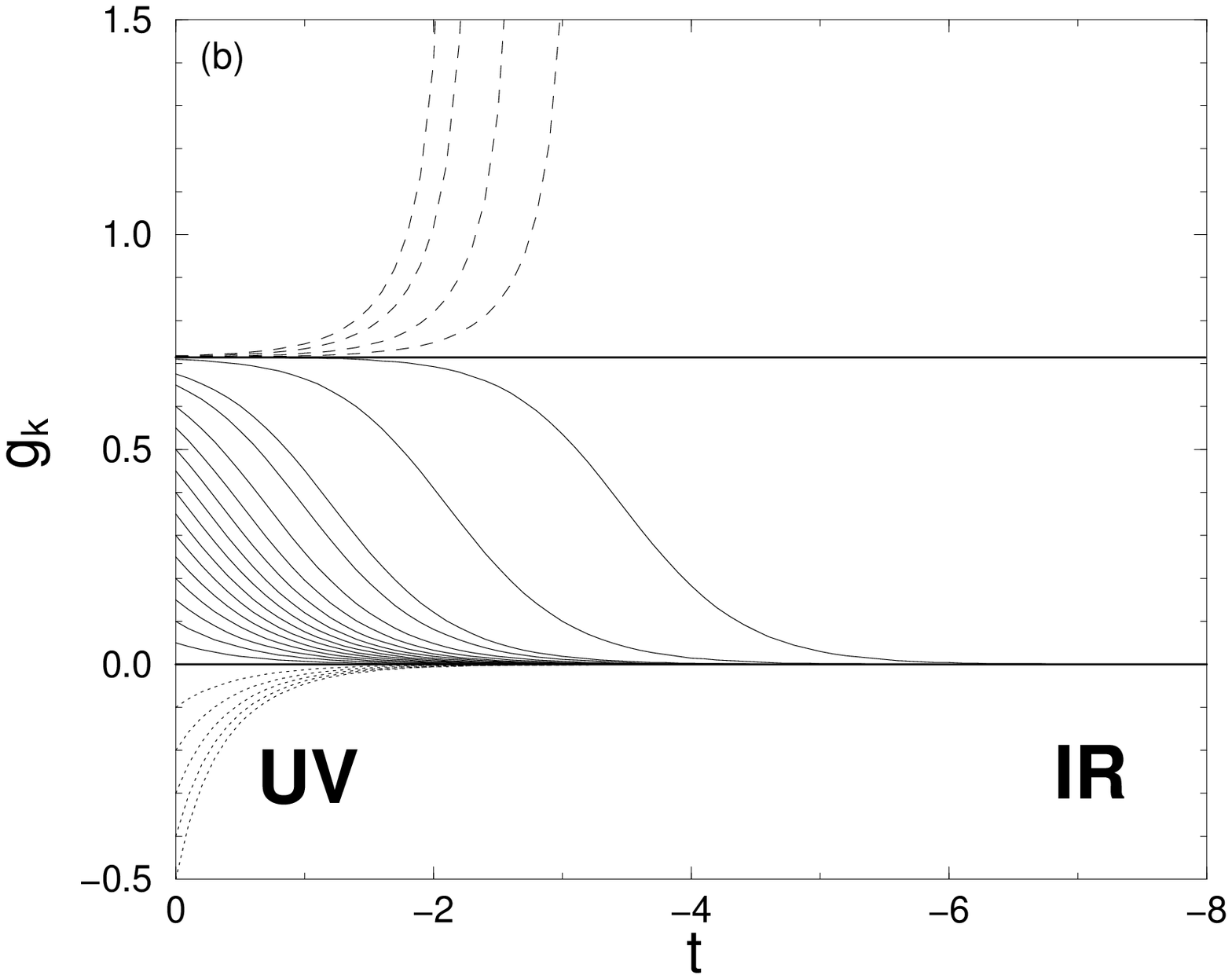}
\end{center}
\parbox{160mm}{\footnotesize
Fig. 1 : The gauge dependence of the UV NGFP of the dimensionless Newton
constant in $d=4$ (a), and the RG flows of it in $\alpha=1$ (b).}
\end{figure}

As immediately recognize, Eq.~(\ref{eq:FPG4}) has the singularity
at $\alpha_{\rm sing}=(\pi^2-114)/54$.
The existence of a singularity is not the problem,
since this point is not the GFP.
In the range $\alpha_{\rm sing}<\alpha$,
$g^*$ has a positive value except for the limit $\alpha\rightarrow\infty$.
On the other hand, in the range $\alpha<\alpha_{\rm sing}$,
$g^*$ has a negative value except for the limit $\alpha\rightarrow-\infty$.
In the limit $\alpha\rightarrow\pm\infty$, the NGFP merges to the GFP.
This problem will be discussed in Sec.~5.
The behavior of Eq.~(\ref{eq:FPG4}) is shown in Fig.~1~(a).
In this figure, a vertical long dashed line corresponds to the
singular point. A solid line is the positive NGFPs,
and a dashed line is the negative NGFPs.
In below, we consider only the positive NGFP. Hence, the shadowed
region in this figure is ignored. 

The typical RG flows are shown in Fig.~1~(b), if
$\alpha$ is fixed to unity.
Under the present approximation given by Eq.~(\ref{eq:RHS2}),
only the FPs are
reliable. However, when $\alpha=1$,
it is possible to study the behavior of
the RG flows \cite{Souma,Reuter}.
In this figure, the horizontal bold solid line, $g^*=0.7152$,
corresponds to the UV NGFP (the circle in Fig.~1~(a)).
This line separates the phase space into two regions;
the strong coupling phase and the weak coupling phase.

\subsection{Cutoff function dependence}
As mentioned previously, constraints for the cutoff function are given by
Eq.~(\ref{eq:cut-const}).
Hence, any functions satisfying these conditions are applicable.
Now, we slightly modify Eq.~(\ref{eq:CUTF}) as
\[
R_1^{(0)}(x,s)=\frac{sx}{\exp(sx)-1},\hspace{5mm}s>0.
\]
Here, $s$ parameterizes the profile of the cutoff function.
As same as the previous subsection, the candidate of the UV NGFP is given
by
\be
g^*=\frac{-2}{B_1(\alpha,s)-2B_2(\alpha,s)}.
\label{eq:FPGl}
\ee
Here, $B_i(\alpha,s),\;(i=1,2)$ is defined similarly to
$B_i(\alpha)$,  except for $\Phi^i_i(0,s)$ and
$\widetilde\Phi^i_i(0,s)$, $(i=1,2)$ depending on $s$.

\begin{figure}[bht]
\epsfxsize=8.5cm
\centerline{\epsfbox{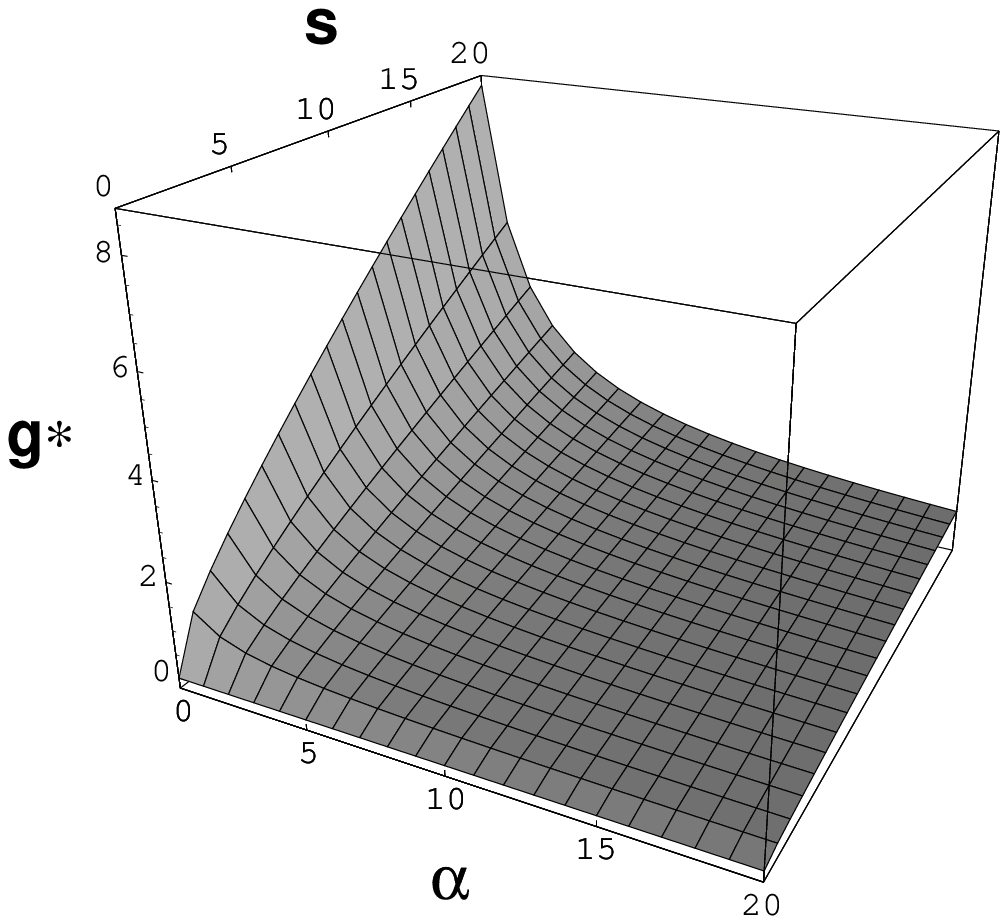}}
\parbox{160mm}{\footnotesize
Fig.2 : The gauge and cutoff function dependence of the UV NGFP of
the dimensionless Newton constant.}
\end{figure}

If we numerically calculate the integral of $\Phi_i^i(0,s)$ and
$\widetilde\Phi^i_i(0,s)$ and insert these into Eq.~(\ref{eq:FPGl}),
we get Fig.~2.
In this figure, the surface represents the position of the UV NGFPs.
The region above the surface is the strong coupling phase,
and that below the surface is the weak coupling phase.
From this figure, it is confirmed that the UV NGFPs exist in a global
range of $\alpha$ and $s$.
The gauge dependence is same as the previous subsection.
For the cutoff function dependence, it is recognized that
the UV NGFPs merge to the GFPs in the limit $s\rightarrow0$.
The reason is expressed as follows. 
In the limit $s\rightarrow0$, the integration of the
cutoff function diverges, therefore, $B_i(\alpha,0)$ function diverges.
Hence, the denominator in Eq.~(\ref{eq:FPGl}) goes to infinity, and
$g^*$ goes to zero.
However, there is no IR cutoff in the limit $s\rightarrow0$.
Hence, this limit is out of the applicability of the ERGE.
If we take another type of the cutoff functions such as $\exp(-sx)$,
same structure is observed.
Therefore, the cutoff function dependence does not cause the
disappearance of the UV NGFP and the change of the phase structure.

\section{Summary and discussion}
\noindent
In this article, we consider the gauge
and cutoff function dependence of the
UV NGFP.
For the cutoff function dependence,
the UV NGFP strongly depends on the profile of the cutoff function.
However, this dependence does not change the phase structure of pure QG.
Though there are many cutoff functions, the phase structure will not
be changed by this dependence.

For the gauge dependence, the UV NGFP exists in a global range of
$\alpha$
except for $\alpha=\pm\infty$.
This gauge, $\alpha=\pm\infty$, is a bad gauge in this formulation.
The disappearance of the UV NGFP for this gauge
is a serious problem. This is because,
the phase structure of pure QG is changed in this gauge.
This problem may be due to the treatment of the
constant gauge parameter.
Hence, if we will be able to treat the gauge parameter as the running gauge
parameter, this problem may disappear.

In this article, only the operators that are invariant under general coordinate
transformations are considered, because the gauge symmetry is
preserved by the projection on $g_\mn=\gb_\mn$.
However, if we improve this approximation to treat the running
gauge parameter, the gauge symmetry 
is not manifestly maintained.
Thus, the operators that can not preserve general coordinate
transformations must be included in the functional space \cite{Gauge}.

The other improvement is
to formulate the ERGE without the gauge fixing.
Recently, for pure $SU(N)$ gauge theory,
Tim R. Morris proposed the formulation of
the gauge invariant ERGE \cite{Morris}.
This formulation does not need the gauge fixing.
Hence, if we can formulate the ERGE for QG in this formulation
and show the existence of the UV NGFP, QG becomes
an asymptotically safe theory and (non-perturbatively)
renormalizable \cite{Souma-p}.

In this article, the Einstein-Hilbert truncation is applied.
However, for these two improvements,
to accurately study the existence of the UV NGFP and the
phase structure, we need the
extension of the functional space. One possibility is to include
the $R^2$-terms. The other is to use the method so called frame
transformations \cite{Magnato}. By this transformations, the higher derivative
gravity is reduced to the Einstein gravity with the auxiliary tensor
matter field. Hence, if we study this reduced theory, the phase
structure of the higher derivative gravity will be clarified.

\section*{Acknowledgements} 
We would like to thank H. Aoyama, M. Sakagami, J. Soda and J-I. Sumi
for useful suggestions and discussions.
The part of numerical computations in this work was carried out 
at the Yukawa Institute Computer Facility.

\end{document}